\documentclass[%
 aip,
 amsmath,amssymb,
 reprint,%
]{revtex4-1}

\usepackage{graphicx}% Include figure files
\usepackage{dcolumn}% Align table columns on decimal point
\usepackage{bm}% bold math
\usepackage[utf8]{inputenc}
\usepackage[T1]{fontenc}
\usepackage{mathptmx}
\usepackage{etoolbox}

\makeatletter
\def\@email#1#2{%
 \endgroup
 \patchcmd{\titleblock@produce}
  {\frontmatter@RRAPformat}
  {\frontmatter@RRAPformat{\produce@RRAP{*#1\href{mailto:#2}{#2}}}\frontmatter@RRAPformat}
  {}{}
}%
\makeatother
\begin{document}

\preprint{AIP/123-QED}

\title[Sample title]{Pulse formation mechanisms switching in hybrid mode-locked fiber laser}
% Force line breaks with \\
\author{Chenyue Lv}

\author{Baole Lu}%
\email{corresponding author: lubaole1123@163.com.}

\author{Jintao Bai}
\email{corresponding author: baijt@nwu.edu.cn.}
\affiliation{ 
	Statel Key Laboratory of Energy Photon-Technology in Western China, International Collaborative Center on Photoelectric Technology and Nano Functional materials, Institute of Photonics \& Photon-technology, Northwest University, Xi’an 710127, China%\\This line break forced with \textbackslash\textbackslash
}%
\affiliation{%
	Shaanxi Engineering Technology Research Center for Solid State Lasers and Application, Provincial Key Laboratory of Photo-electronic Technology, Northwest University, Xi’an 710069, China%\\This line break forced% with \\
}%

\date{\today}% It is always \today, today,

\begin{abstract}
	Hybrid mode-locking has been widely used in enhancing pulse quality, however how the hybrid two mode-locking techniques work remains unclear. In this paper, we experimentally investigate three pulse formation mechanisms in saturable absorbers (SA) and nonlinear polarization evolution (NPE) passively hybrid mode-locked fiber laser, which are SA-dominated, NPE-dominated, and SA-NPE co-domination switching. Clarified the exists dynamic competition and cooperation among the mode-locking techniques. For the first time, the method of simulating filtered gain spectrum with customized filtering is proposed, and the switching of pulse formation mechanisms is numerically investigated using the coupled Ginzburg-Landau equations. Our results deepen the understanding of hybrid mode-locked fiber lasers and provided a foundation for multi-wavelength mode-locked lasers with different mode-locking techniques in a single cavity.
\end{abstract}

\maketitle

\section{Introduction}

Passively mode-locked fiber lasers are characterized by narrow pulse width, high peak power, compact structure, and low cost. They are widely applied in micromachining \cite{1}, biomedicine \cite{2}, spectral analysis \cite{3}, and optical communication \cite{4}, which is one of the frontier hot research directions in the ultrafast laser field. Since the fiber core is in the micron scale, abundant nonlinear phenomena occur when the power is increased \cite{5}, providing an ideal platform for studying various pulse evolution and soliton dynamics. After years of research, many different passively mode-locking techniques have been developed, including low-dimensional nanomaterial saturable absorbers (SAs) \cite{6}, nonlinear optical/amplifying loop mirrors (NOLM/ NALM) \cite{7,8}, nonlinear polarization evolution (NPE) \cite{9}, semiconductor saturable absorber mirrors (SESAMs) \cite{10}, nonlinear multimode interference (MMI) \cite{11}, and Mamyshev oscillators \cite{12}, etc. These techniques are widely used in mode-locked fiber lasers, but each of them inevitably has some disadvantages that are difficult to overcome. For example, the NPE mode-locking technology has outstanding advantages in output pulse performance but has poor stability and is susceptible to loss of mode lock due to environmental disturbances compared to SA; the SA mode-locking technology is easy to achieve self-starting but has a low damage threshold.

To further address the shortcomings caused by the mode-locking technique and improve laser performance, an idea has been proposed to incorporate other mode-locking techniques to complement the weaknesses of the single mode-locking technique. This combination of more than two mode-locking technologies is called hybrid mode-locking, which has attracted a lot of interest from researchers since its introduction. Experimentally, a hybrid mode-locking technology combining NPE and SESAM mechanisms was proposed by L. Lefort et al. to improve laser performance, such as enhanced cavity stability and narrowed pulse width \cite{13}. Bo Xu et al. developed a hybrid mode-locked device consisting of SA and NALM, which reduced the mode-locking threshold from 330 $mW$ to 120 $mW$ with excellent stability by adding SA \cite{14}. Maria Chernysheva et al. combined SA and NPE mode-locking techniques to observe the generation of multiple soliton complexes in mode-locked Thulium-doped fiber laser, in which SA guarantees high modulation depth, fast relaxation, and NPE ensures multiple solitons structure generation \cite{15}. Yanyan Zhang et al. used NALM and NPE mode-locking techniques to study optical frequency combs. The experimental results show that the hybrid mode-locked fiber laser possesses low relative intensity noise characteristics based on NPE and self-starting characteristics based on NALM \cite{16}. Yong Yao et al. combined the MMI and NPE mode-locking techniques in multimode fibers to build a hybrid mode-locked laser and experimentally obtained a bound-state pulse consisting of more than 30 sub-pulses with switchable center wavelength covering the entire C band \cite{17}. It seems that all experimental results indicate that the two mode-locking techniques cooperate intimately to enhance the cavity properties. However, Zheng Zheng et al. demonstrated a unidirectional dual-laser comb source using NPE and SA mode-locking, with two different pulse states corresponding to different mode-locking mechanisms \cite{18}. Theoretically, some numerical models based on the coupled Ginzburg-Landau equations (CGLE) were used for the analysis of hybrid mode-locking fiber lasers \cite{19,20,21,22}. In the previous studies, most of the numerical results were positive for hybrid mode-locking. In other words, most of the studies showed that the two mode-locking techniques positively synergize in the cavity to improve mode-locking laser quality. Until recently, Li Li et al. found that the NPE technique negatively affects linear absorption or even reverse saturation absorption on the pulses. For an NPE and SA hybrid mode-locked laser, there is a dynamic synergy between two mode-locking techniques, rather than a static combined mechanism \cite{23}. Therefore, the multiple mode-locking techniques in hybrid mode-locked lasers are not as harmonious as they appear, both theoretically and experimentally, and further research is still needed to investigate how the mode-locking techniques in hybrid mode-locked lasers work.

In this study, we investigated the switching process of the pulse formation mechanism in SA and NPE passively hybrid mode-locked Erbium (Er)-doped fiber lasers. The mode-locked pulses are observed in three cases: dominated by SA technology, NPE technology, and cooperation by SA and NPE, explain the mechanism of generating two sets of pulses in the same laser cavity based on different mode-locking techniques. It is demonstrated that there is dynamic competition and cooperation between the mode-locking mechanisms in a hybrid mode-locked fiber laser. The pulse generation is strongly correlated with the strong side of the mode-locking mechanism, while the other weaker mode-locking mechanism will act as an auxiliary. A set of coupled Ginzburg-Landau equations is theoretically constructed, and for the first time, we propose to simulate the filtered gain spectrum with a customized filter. The numerical simulation results are in good agreement with the experimental results. Our results induce reflection on the hybrid mode-locking mechanism and deepen the understanding of hybrid mode-locked fiber lasers. 

\section{Experimental Results and Discussion}

The experimental configuration of the all polarization maintaining (PM) hybrid mode-locked Er-doped fiber laser is shown in Fig. 1. About 2.15 $m$ Er-doped fiber (EDF, PM-ESF-7/125) is pumped by a stabilized 980 $nm$ laser diode (LD) via a 980/1550 $nm$ wavelength division multiplexer (WDM). The laser operates in a hybrid mode-locking state of NPE and SA (carboxyl-functionalized graphene oxide, GO-GOOH) \cite{6}. The fiber structure to generate the NPE effect consists of three rotating segments of polarization maintaining fiber (PMF) with a length ratio of 1:2:1, a splicing angle of $90^{\circ}$, an input angle of $30^{\circ}$, and an output angle of $45^{\circ}$ \cite{9}. A polarization controller (PC) is placed in the second fiber section of the NPE structure to accurately regulate the polarization state of the laser. In addition, an PM-isolator (PM-ISO) is used to ensure unidirectional laser transmission within the cavity; a 10:90 optical coupler (OC) is used with $90\%$ of the ports connected into the cavity and $10\%$ of the ports for pulse output. All devices and fibers used in the cavity are of PM construction and are connected by $\sim$ 11.5 $m$ PMFs. The group velocity dispersion (GVD) values for the EDF and PMF used in the laser are $\sim$ -22 $ps^{2}/km$ and $\sim$ -20 $ps^{2}/km$ at 1550 $nm$, respectively. The total cavity length is $\sim$ 13.65 $m$, which gives a calculated net cavity dispersion estimated to be -0.296 $ps^{2}$ in the anomalous dispersion region. For output pulse measurement, a high-speed real-time oscilloscope (OSC, Agilent, DSO9104A) is used to track temporal information, an optical spectrum analyzer (OSA, Yokogawa, AQ6370C) to record time-averaged optical spectrum, a radio frequency (RF) spectrum analyzer (Keysight, N9000B CXA signal analyzer) for RF spectrum, and an autocorrelator (APE, PulseCheck-50) to measure the width of pulses.

\begin{figure}
	\includegraphics[width=6.8cm]{Laser}
	\caption{\label{fig:epsart}Configuration of all-PM hybrid mode-locked Er-doped fiber laser.}
\end{figure}

The mode-locking threshold for the laser shown in Fig. 1 is about 48 $mW$, but this value is obtained by gradually decreasing the pump power after mode-locking stabilization. Increasing the pump power to 60 $mW$, three pulse formation mechanisms based on SA domination, NPE domination, and SA-NPE co-domination can be observed in the all-PM hybrid mode-locked fiber laser. These mechanisms can be switched by simply adjusting the PC, and the pulse characteristics under those three are shown in Fig. 2.

\noindent\textbf{SA domination.} When the pump power reached 60 $mW$, pulses based on the SA mode-locking mechanism were obtained without deliberate adjustment of the PC. The pulse train in the SA mode-locked domination is shown in Fig. 2(a), from which the splitting of the pulse can be observed, and the generated pulse is near the middle of the original. Figure 2(b) illustrates the time-averaged optical spectrum corresponding to the pulse train. The spectrum has a 3 $dB$ bandwidth of 2.58 $nm$ at the center wavelength of 1562.59 $nm$ and low sidebands. Compared with the conventional soliton spectrum produced by typical SA mode-locking, we attribute the narrower spectral width and lower sideband generation to pulse splitting. Pulse splitting results in low individual pulse energy, which are insufficient to support further spectral broadening and the creation of more Kelly sidebands. Figure 2(c) displays the RF spectrum, which exhibits a high signal-to-noise ratio of 70.6 $dB$ at the fundamental frequency of 29.88 $MHz$. The inset shows the RF spectrum train in the range of 0 to 1 $GHz$. Although there is still a 14.94 $MHz$ signal residual in the RF spectrum, it is much closer to the second harmonic overall, which is why we present 29.88 $MHz$ as the fundamental frequency. The autocorrelation (AC) trace of the mode-locked pulse and the fitting curve based on the $Sech^{2}$ function are shown in Fig. 2(d). The pulse width is calculated to be 1156 $fs$. Combined with the spectrum, a time-bandwidth product (TBP) of 0.3662 is obtained. The TBP is slightly larger than the limit, which indicates a certain chirp of the pulse. The pulse self-starting without tuning the PC, which means that the SA technique dominates the mode-locking at this point, while the NPE effect serves as a secondary.
\begin{figure*}
	\includegraphics[width=16cm]{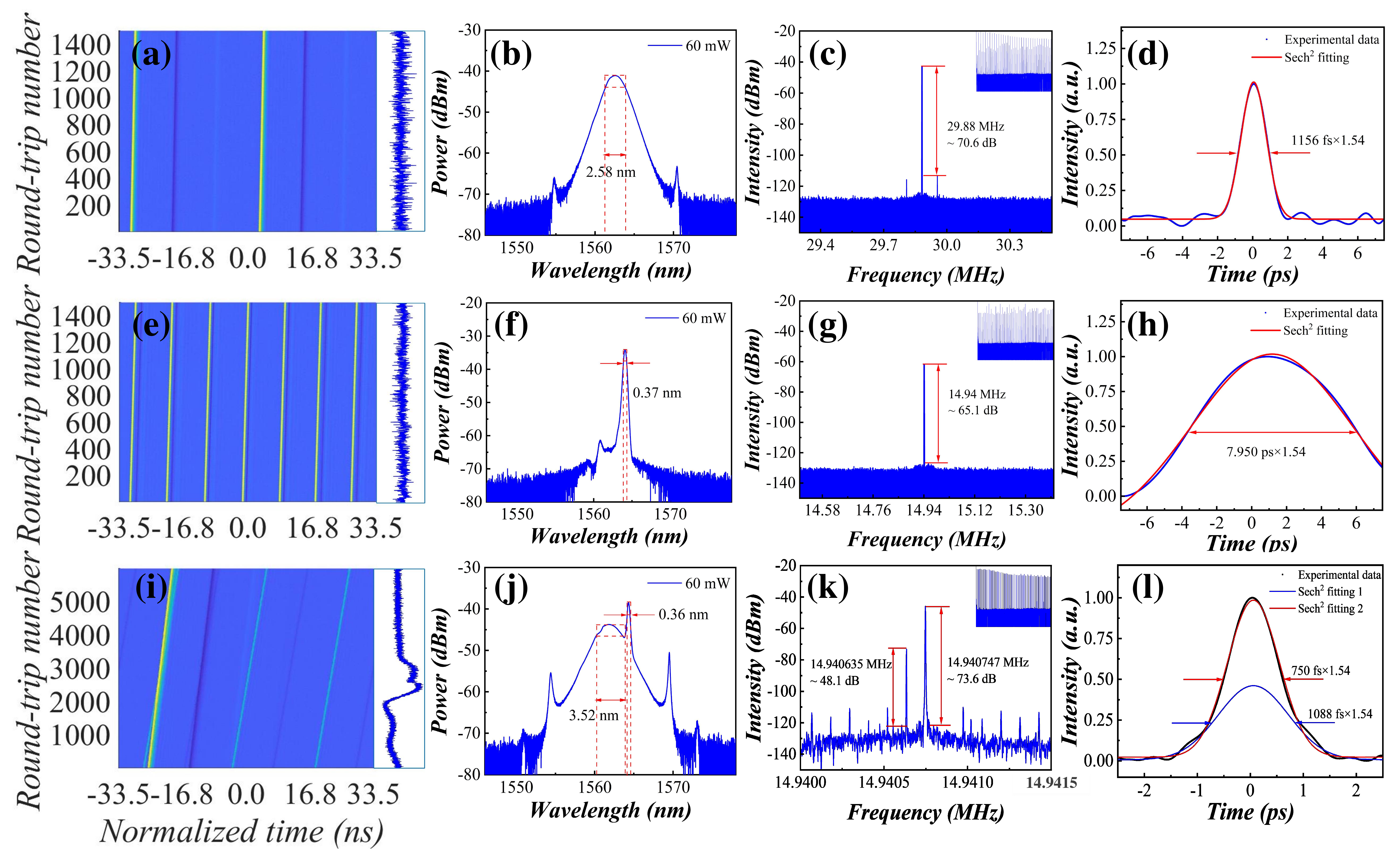}
	\caption{\label{fig:wide}Output characteristics of the all-PM hybrid mode-locked fiber lasers. (a)-(d) SA domination. (e)-(h) NPE domination. (i)-(l) SA-NPE co-domination.}
\end{figure*}
\noindent\textbf{NPE domination.} Carefully adjusting the PC brings the polarization state in the cavity to a specific point, and the intensity of the mode-locking pulse suddenly decreases and splits into more sub-pulses. The pulse trains, time-averaged optical spectrum, RF spectrum, and AC traces obtained at this point are shown in Fig. 2(e)-(h). We can obtain that the pulse has a 3 $dB$ bandwidth of 0.37 $nm$ at the center wavelength of 1564.10 $nm$. Compared to SA domination, the spectrum compression is nearly seven times narrower. In the RF spectrum, the fundamental frequency has a high signal-to-noise ratio of 65.1 $dB$ at 14.94 $MHz$, but the long-range RF spectrum train is not neat due to the interference caused by multiple pulse splitting. The pulse width of 7.95 $ps$ is obtained after $Sech^{2}$ fitting, corresponding to a TBP of 0.3605, slightly chirped. Taking into account the fact that the spectrum bandwidth of the SA mode-locking is not easily affected to produce such a large change, as well as the filtering characteristics of the NPE mode-locking. We believe that the dominant mode-locking mechanism in the cavity at this time has been changed, and the pulse is generated by the NPE mode-locking technology, while the SA technology is used as an auxiliary. That also means that the NPE in the cavity is stronger than the SA in this particular polarization state.

\noindent\textbf{SA-NPE co-domination.} Constantly adjusting the PC, we achieved the coexistence of the two pulses described above. As shown in Fig. 2(i), the pulse train exhibits one large pulse with three small pulses. During continuous measurements of about 6000 circles, an interference of intensity can be observed which is produced by the intersection of large and small pulses. With the spectrum in Fig. 2(j), we can observe dual-wavelength pulses, i.e., the pulses previously guided by both SA and NPE mode-locking techniques both preserved well. The center wavelengths of the two sets of pulses are 1562.14 $nm$ and 1564.32 $nm$, and their corresponding 3 $dB$ bandwidths are 3.52 $nm$ and 0.36 $nm$, respectively. The RF spectrum in Fig. 2(k) exhibits two independent fundamental frequencies, 14.940747 $MHz$ vs. 14.940635 $MHz$, and their difference in repetition frequency is 112 $Hz$, corresponding to the wavelength difference.  Figure 2(l) shows the corresponding AC traces and the two-pulse $Sech^{2}$ fitting curves, where the narrower pulse corresponds to the SA and the wider pulse in the base bulge corresponds to the NPE mode-locking. It should be pointed out that since the pulses themselves have different group velocities, the AC trace bump is only able to determine the simultaneous presence of both pulses. With the AC trace, it is possible to fit accurately the pulse width of the stronger one as 760 $fs$, which corresponds to a TBP of 0.3243 with a slight chirp. For the other set of weaker pulses, as evidenced by the TBP obtained according to the fit is 0.048 \textemdash\ much smaller than the transform limit, the AC trace could not be fitted to accurately determine the pulse width.
\begin{figure}
	\centering\includegraphics[width=8cm]{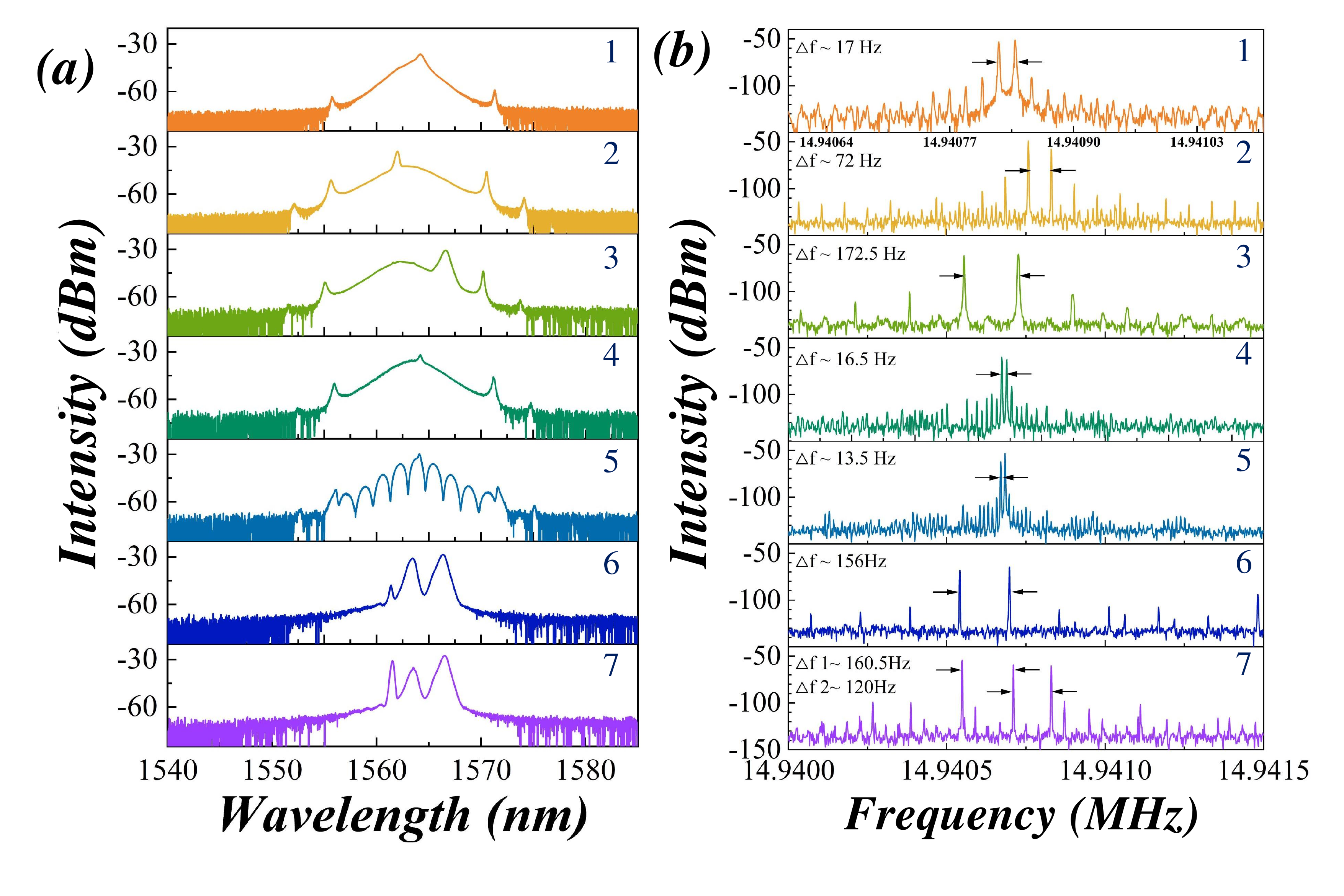}
	\caption{Schematic of hybrid mode-locked dual-wavelength switching. 
		(a) Time-averaged optical spectrum. (b) RF spectrum.}
\end{figure}
Slightly increasing the pump power to 61 $mW$ and carefully tuning the PC results in switchable, narrowly tunable dual- and triple-wavelengths with more repetitive frequency differences. These results are reflected in Fig. 3. After computational analysis, it is found that the wavelength of the mode-locking center dominated by SA is determined to fluctuate around 1563 $nm$ with a range of less than 2 $nm$. In contrast, the mode-locked pulse dominated by NPE fluctuates around three frequency bands 1561 $nm$, 1564 $nm$, and 1566 $nm$. The sixth and seventh subplots of Fig. 3 show a rather strange intensity distribution, with the center wavelengths of the three peaks in the three pulses that appear briefly being 1561.54 $nm$, 1563.52 $nm$, and 1566.54 $nm$, respectively. Considering the wavelength distributions of the SA versus NPE domination pulses, we believe that the middle peaks are still SA domination mode-locked, although the energy is robbed by the NPE, not enough to produce obvious sidebands. This triple-wavelength phenomenon is considered to be the gradual approach of the intracavity hybrid mode-locking effect to NPR dominance, which enhances the composite filtering effect and leads to the appearance of two other pulses on both sides of the SA-dominated mode-locking pulse.

Combined with all these experimental phenomena, there is sufficient evidence to believe that SA and NPE in the cavity have a competitive relationship in addition to the synergistic effect. When there is an overwhelming dominance of one mode-locking technique in the cavity, the other mode of mode-locking exists only as an auxiliary. If neither SA nor NPE is overwhelmingly dominant, this competition manifests itself, resulting in a dynamic exchange of energy between the two mode-locking methods, leading to the emergence of this dual-wavelength pulse train guided by different mode-locking techniques.

The reason why the relative strength magnitude of SA and NPE can be changed by simply adjusting the PC in the laser cavity shown in Fig. 1, is due to the special characteristics of all-PM NPE structures \cite{9}. There is a large birefringence in a PM fiber, so when two orthogonally polarized beams are transmitted in a PM fiber, they will gradually move away from each other and generate group velocity mismatch (GVM). To compensate for the GVM, all-PM NPE structure needs to be rotating spliced. The PC on the second fiber of the spliced structure can control the polarization state in the all-PM cavity to a certain extent, thus affecting the amount of GVM and achieving the purpose of controlling the strength of the NPE effect.

\section{Numerical Simulation}

Considering that the theoretical study should be universal, for the hybrid all-PM Er-doped mode-locked fiber laser shown in Fig. 1, we have carried out a structural simplification. The theoretical model goes more focused on its properties as a hybrid mode-locked fiber laser rather than being limited to the PM structure. We have developed a theory based on the coupled Ginzburg-Landau equations (CGLE) and performed numerical simulations. The equations are shown below:

\begin{widetext}
\begin{equation}
\begin{aligned}
		\frac{\partial \varphi_{x}}{\partial z}& = -i \frac{\beta_{2}}{2} \frac{\partial^{2} \varphi_{x}}{\partial t^{2}} + i\gamma \left(\left|\varphi_{x}\right|^{2}+\frac{2}{3}\left|\varphi_{y}\right|^{2}\right) \varphi_{x}+\frac{i \gamma}{3} \varphi_{x}^{*} \varphi_{y}^{2}+\frac{g-l}{2} \varphi_{x}+\frac{g}{2 \Omega_{g}^{2}} \frac{\partial^{2} \varphi_{x}}{\partial t^{2}}\\
		\frac{\partial \varphi_{y}}{\partial z}&= -i \frac{\beta_{2}}{2} \frac{\partial^{2} \varphi_{y}}{\partial t^{2}} +i \gamma \left(\left|\varphi_{y}\right|^{2}+\frac{2}{3}\left|\varphi_{x}\right|^{2}\right) \varphi_{y}+\frac{i \gamma}{3} \varphi_{y}^{*} \varphi_{x}^{2}+\frac{g-l}{2} \varphi_{y}+\frac{g}{2 \Omega_{g}^{2}} \frac{\partial^{2} \varphi_{y}}{\partial t^{2}}\\
\end{aligned}
\end{equation}
\end{widetext}

\noindent where $\beta_{2}$ is GVD; $\gamma$ is the nonlinear coefficient and $l$ is the loss coefficient of the fiber;  $\Omega _{g}$ is the gain bandwidth; the gain coefficient in an EDF can be described by the following equation:
\begin{equation}
	g=g_{0} \exp \left(-\frac{\int\left(\left|\varphi_{x}\right|^{2}+\left|\varphi_{y}\right|^{2}\right) d t}{P_{ sat }}\right)
\end{equation}
\noindent where $g_{0}$ is the small signal gain and $P_{ sat }$ is the gain saturation energy. In addition, a Gaussian-type bandpass filter with two wavelengths and two pulse widths is used to simulate the composite filter model generated by the superposition of gain filtering, birefringent filtering, and NPE filtering.
\begin{equation}
	T_{f}(\lambda)=A_{1} \exp \left(-\frac{\left(\lambda-\lambda_{1}\right)^{2}}{2 \Delta \lambda_{1}^{2}}\right)+A_{2} \exp \left(-\frac{\left(\lambda-\lambda_{2}\right)^{2}}{2 \Delta \lambda_{2}^{2}}\right)	
\end{equation}
\noindent where $A_{1}$, $A_{2}$ are the intensities of the two Gaussian peaks; $\lambda_{1}$, $\lambda_{1}$, $\Delta \lambda_{1}$ and $\Delta \lambda_{2}$ are the center wavelengths and bandwidths of the two Gaussian peaks.

The transmittance of SA is given by $T=1-\alpha_{ns}-\alpha_{0} \exp \left(-\left|\varphi_{x,y}\right|^{2} / P_{\text {sat}}\right)$, where  $\alpha_{ns}$, $\alpha_{0}$, $P_{sat}$ are the non-saturable loss, the modulation depth, and the saturation power of SA. In the hybrid mode-locking scheme, $\alpha_{ns}$, $\alpha_{0}$, and $P_{sat}$ were set as 0.1, 0.097, and 30 $W$ \cite{6}, respectively. As for the transmission matrix model of the NPE, unlike the complex model of the all-PM NPE structure that we have used before \cite{24}, the more concise but common model of dual PCs with polarization dependent isolator (PD-ISO) is adopted for numerical simulation.
\begin{equation}
	\left[\begin{array}{l}
		\varphi_{outx } \\
		\varphi_{outy }
	\end{array}\right]=J_{P C 2} J_{P D-I S O} J_{P C 1}\left[\begin{array}{l}
		\varphi_{inx} \\
		\varphi_{iny}
	\end{array}\right]
\end{equation}
\noindent where the transmission matrix for PC is:
\begin{equation}
	J_{PC}=\left[\begin{array}{cc}
		\cos (\alpha) & -\sin (\alpha) \\
		\sin (\alpha) & \cos (\alpha)
	\end{array}\right]\left[\begin{array}{cc}
		e^{i \phi / 2} & 0 \\
		0 & e^{i\phi / 2}
	\end{array}\right]
\end{equation}
\noindent and the transmission matrix for PD-ISO is:
\begin{equation}
	J_{P D-I S O}=\left[\begin{array}{cc}
		\cos (\beta) & -\sin (\beta) \\
		\sin (\beta) & \cos (\beta)
	\end{array}\right]\left[\begin{array}{cc}
		1 & 0 \\
		0 & 0
	\end{array}\right]	
\end{equation}
\noindent where $\alpha$, $\beta$ is the deflection angle from the PC and PD-ISO; $\phi$ is the phase delay introduced in the cavity. The parameters used in the numerical simulations correspond to the fiber parameters used in the experiments shown in Table 1. 
\begin{table*}
\begin{ruledtabular}
		\centering
		\caption{Parameters in the numerical simulation.}
		\begin{tabular}{cccccc}
			Variable name &Value & Variable name & Value& Variable name & Value \\
			\hline
			$\beta_{2PMF}$ & $-22\: ps^{2}/km$ & $\beta_{2EDF}$ & $-20\: ps^{2}/km$ & $\gamma_{PMF}$ & $1.3 \: W^{-1}km^{-1}$  \\
			$\gamma_{EDF}$ & $4.7\: W^{-1}km^{-1}$ & $g_{0}$ & $27.68$ & $P_{sat}$ & $30\: W$\\
			$l$ & $0.15\: m^{-1}$ & $\Omega_{g}$ & $40\: nm$ & $A_{1}$ & $1$\\
			$A_{2} $ & $Varying$ & $\lambda_{1}$ & $1560\: nm$ & $\lambda_{2}$ & $1565\: nm$\\
			$\Delta \lambda_{1}$ & $20\: nm$ & $\Delta \lambda_{2}$ & $4\: nm$ & $\alpha_{PC1}$ & $100^{\circ}$\\
			$\beta$ & $100^{\circ}$ & $\alpha_{PC2}$ & $30^{\circ}$\\
	\end{tabular}
	\end{ruledtabular}
	\label{tab:shape-functions}
\end{table*}
The intracavity filtering state is changed by varying the value of $A_{2}$, in order to simulate the change in filtering intensity brought about by the change in intensity of the intracavity NPE effect. By solving the above model by Matlab using the fourth-order Runge-Kutta methods, we obtain three locking dominant processes in SA and NPE hybrid mode-locked fiber lasers corresponding to the experimental results as shown in Fig. 4. When $A_{2} = 1/20$, the SA domination mode-locking pulse is obtained, and its temporal evolutions and optical spectrum evolutions during the 1500 pulse round trip, and field autocorrelation of the last turn are represented in Fig. 4(a)-(c), respectively. The original input signal is a Gaussian-type noise signal, and as the number of round trips increases SA dominates and rapidly develops into a stable mode-locking. Being a mode-locking pulse dominated by SA, it has significant sidebands, a wide bandwidth, and a narrow pulse width, as observed in our experiments. Fig. 4(d)-(f) show the NPE domination pulse generated at $A_{2}= 1/3$, which can be seen from the temporal evolutions to drift significantly to the right with the existing simulation window, implying that the generated pulse has a different fundamental frequency. Under the same small-signal gain conditions, the NPE domination pulse is more significantly affected by its filtering effect, resulting in being limited to a narrow bandwidth, while causing a reduction in the tolerable energy of the pulse and splitting it. It can also be seen from the autocorrelation trace in Fig. 4(f) that the NPE dominated mode-locked pulse width is significantly wider.
\begin{figure*}
	\centering\includegraphics[width=12cm]{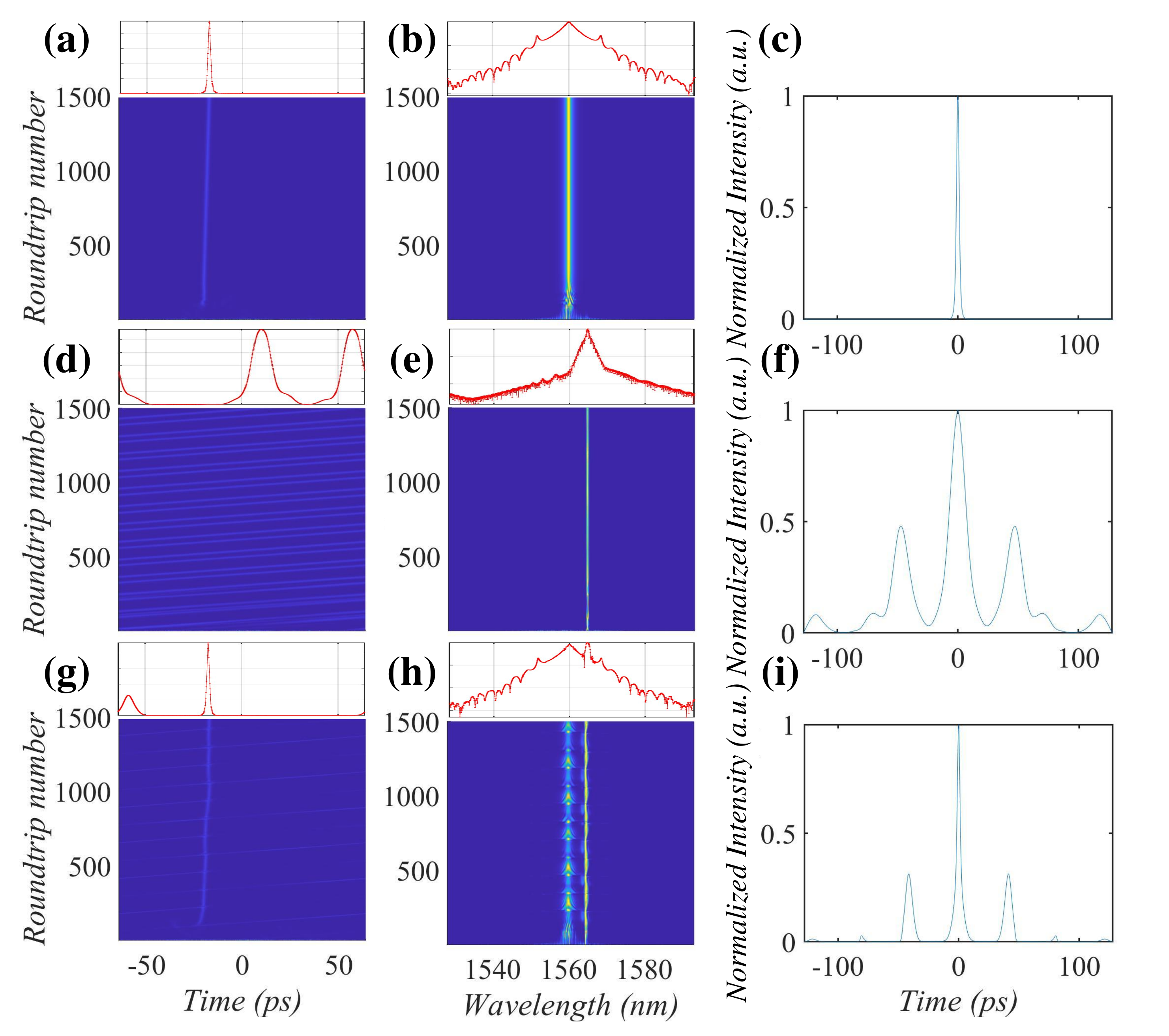}
	\caption{Numerical simulation results of the hybrid mode-locked fiber lasers.(a)-(c) SA domination. (d)-(f) NPE domination. (g)-(i) SA-NPE co-domination.}
\end{figure*}
After adjusting $A_{2}$ to $1/12.778$, the two mode-locking techniques’ strengths reach an approximate agreement, at which time the mode-locking state becomes SA-NPE co-domination, as presented in Fig. 4(g)-(i). The temporal evolutions of the pulse for a 1500-turn round trip in Fig. 4(g) illustrate that there are two sets of coexisting mode-locking pulses in the cavity. During the transmission process, two sets of pulses collide in the cavity at a constant period due to different group velocity. The collision process generates energy exchange, and the coexistence state of the pulses in the cavity is the result of dynamic energy balance. The optical spectrum evolution process with the final round trip spectra is provided in Fig. 4(h). The shape and center wavelength show that the output spectrum is a superposition of spectrum from two mode-locked techniques, SA and NPE. The AC trace is represented in Fig. 4(i) as a double-pulse AC trace for narrow and wide pulses. In comparison with the AC traces in Fig. 4(c) and (f), the highest centered pulse is the narrower top combined with the wider base, and there should be periodic satellite peaks around the central peak. The results given in the simulation are the real-time AC trace for the last round trip, and the satellite summits are erased when the results are captured with the average autocorrelator. Although we did not observe the presence of satellite peaks due to instrumentation limitations, the shape of the main peak is highly consistent with the results obtained in our experiments. This numerical result, which is in high coincidence with the experiment, also provides sideways evidence of the credibility of the theoretical use of filter strength to characterize the NPE strength.
\begin{figure}[h!]
	\centering\includegraphics[width=8cm]{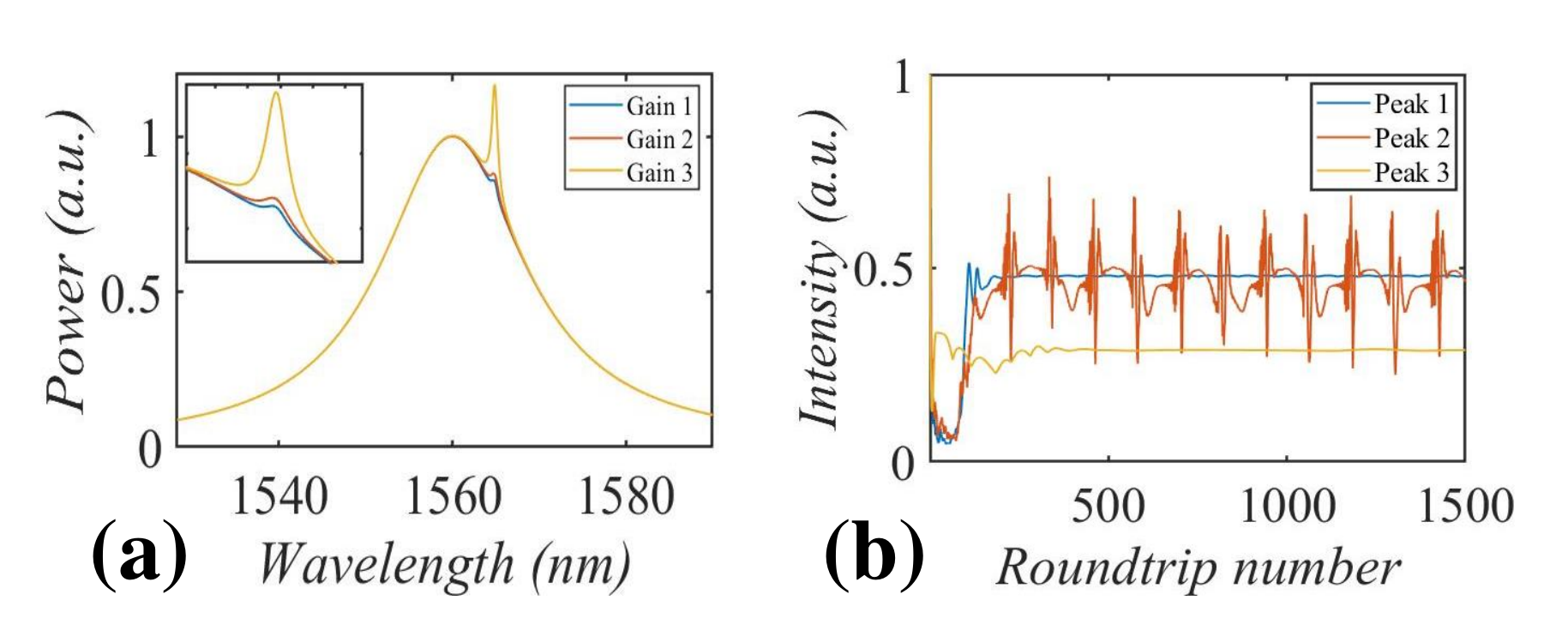}
	\caption{(a) Filtered gain spectrum and (b) intensity versus roundtrip number.}
\end{figure}
The filtered gain spectrum and the variation curve of pulse intensity with the number of round trips are given in Fig. 5, which visualizes the effect of filter intensity on the gain spectrum of the cavity under three different mode-locking dominant conditions as well as the energy change of the pulse evolution process. The inset in Fig. 5(a) magnified shows the variation of the gain spectrum intensity with the narrow filter intensity. In a hybrid mode-locked laser, the blue filtrated gain spectrum produces SA domination mode-locking, the red filtrated gain spectrum produces SA-NPE co-domination mode-locking, and the yellow filtrated gain spectrum produces NPE domination mode-locking. Figure 5(b) illustrates the peak intensity variation of the pulse. The smooth blue and yellow lines are stable mode-locked pulses dominated by SA and NPE, respectively. The yellow line has a lower intensity due to pulse splitting caused by the narrow bandwidth. The red line represent the mode-locked pulse set co-dominated by SA and NPE, and they had 11 collisions in 1500 round trip periods. The reason for such a high frequency of collisions compared to the experimental results where only one collision occurs in nearly 6000 round trips is that the time window used in the simulation is shorter on the order of $ps$ ($ns$ in the experiments), causing the pulses to meet each other after a much shorter number of round trips.

\section{Conclusion}
We experimentally and theoretically investigated the switching process of two pulse formation mechanisms in SA and NPE passively hybrid mode-locked fiber lasers. Experimentally, pulses were observed in three mode-locking technology-dominated cases: SA domination, NPE domination, and SA-NPE co-domination. The experimental results demonstrate the existence of dynamic competition and cooperation in the hybrid mode-locking mechanism, as well as the strong correlation between the final generated pulses and the strong parties in the mode-locking mechanism in hybrid mode-locking. Theoretically, a set of CGLE based on the more general hybrid mode-locked fiber laser is developed and numerically solved using the fourth-order Runge-Kutta method, yielding three different pulse evolution results. The simulation results match the experiments almost perfectly, demonstrating the feasibility of our proposal to use a customized filter to simulate the experimental gain filtered spectrum and to reflect the intensity of the two mode-locking techniques in the hybrid mode-locking technique. Our results have triggered a rethinking of the hybrid mode-locking mechanism, deepen the understanding of the mode-locking mechanism of hybrid mode-locked fiber lasers as well as provided a foundation for the generation of multi-wavelength lasers using different mode-locking techniques.

\begin{acknowledgments}
	This work was supported by National Major Scientific Research Instrument Development Project of China (51927804); Shaanxi Key Science and Technology Innovation Team Project (2023-CX-TD-06).
\end{acknowledgments}

\section*{Author Declarations}
	The authors declare no conflicts of interest.

\bibliography{Ref}% Produces the bibliography via BibTeX.

\end{document}